\documentclass[12pt,onecolumn,twoside,a4paper,draftclsnofoot]{IEEEtran}

\usepackage{amsmath}
\usepackage{mathrsfs}

\usepackage{amsfonts}
\usepackage{cite}
\usepackage{graphicx,color,overpic}
\usepackage{amsmath}
\usepackage{times}
\usepackage{latexsym}
\usepackage{bm}
\usepackage{amssymb}
\usepackage{cases}
\usepackage{array}
\usepackage{fancyhdr}
\usepackage{setspace}
\usepackage{subfigure}
\usepackage{url}
\usepackage{algorithm}
\usepackage{algorithmic}
\usepackage{multirow}

\usepackage{epsfig}
\usepackage{fancybox}
\usepackage{textcomp}
\usepackage{multirow}
\usepackage{setspace}
\usepackage{psfrag}
\usepackage{booktabs}
\usepackage{colortbl}
\usepackage[table]{xcolor}
\usepackage{tabu}

\usepackage{enumerate}

\definecolor{mygray}{gray}{.9}
\definecolor{mypink}{rgb}{.99,.91,.95}
\definecolor{mycyan}{cmyk}{.3,0,0,0}

\newcommand{\tabincell}[2]{\begin{tabular}{@{}#1@{}}#2\end{tabular}}

\begin{document}

\title{\huge{RaPro: A Novel 5G Rapid Prototyping System Architecture}}

%\author{\IEEEauthorblockN{Jun Zhang$^{\S}$, Chau Yuen$^{\S}$, Chao-Kai Wen$^{\dag}$, Shi Jin$^{\ddag}$, and Xiqi Gao$^{\ddag}$}
%\IEEEauthorblockA{$^{\S}$ Singapore University of Technology and Design, Singapore,
%Email: \{zhang\_jun, yuenchau\}@sutd.edu.sg\\}
%\IEEEauthorblockA{$^{\dag}$ Institute of Communications Engineering, National Sun Yat-sen University\\
%Kaohsiung 804, Taiwan, Email: \{ckwen\}@ieee.org\\}
%\IEEEauthorblockA{$^{\ddag}$ National Mobile Communications Research Laboratory, Southeast University\\
%Nanjing 210096, P. R. China, Email: \{jinshi, xqgao\}@seu.edu.cn}\\
%}

\author{\IEEEauthorblockN{Xi Yang, Zhichao Huang, Bin Han, Senjie Zhang, Chao-Kai Wen, \\Feifei~Gao, and Shi Jin}
\thanks{X. Yang, Z. Huang, B. Han and S. Jin are with the National Mobile Communications Research
Laboratory, Southeast University, Nanjing 210096, China (e-mail: \{ouyangxi, huangzhichao, seuhanbi, jinshi\}@seu.edu.cn).}
\thanks{S. Zhang is with Beijing University of Posts and Telecommunications, Beijing 100876, China (e-mail:senjie.zhang@hotmail.com).}
\thanks{C.-K. Wen is with the Institute of Communications Engineering, National
Sun Yat-sen University, Kaohsiung 80424, Taiwan (e-mail: ckwen@ieee.org).}
\thanks{F. Gao is with State Key Laboratory of Intelligent Technology and Systems, Tsinghua National Laboratory for Information Science and Technology, Department of Automation, Tsinghua University, Beijing 100084, China (e-mail:feifeigao@ieee.org).}\vspace{-0.6cm}}

%\author{\IEEEauthorblockN{Xi Yang, Zhichao Huang, Bin Han,~\IEEEmembership{Student Member,~IEEE}, Senjie Zhang, Chao-Kai Wen,~\IEEEmembership{Member,~IEEE}, Feifei~Gao,~\IEEEmembership{Senior~Member,~IEEE} and Shi Jin,~\IEEEmembership{Member,~IEEE}}}

%\author{\IEEEauthorblockN{Jun Zhang$^{\S}$, Chau Yuen$^{\S}$, Chao-Kai Wen$^{\dag}$, Shi Jin$^{\ddag}$, and Xiqi Gao$^{\ddag}$}\\
%
%\IEEEauthorblockA{$^{\S}$ Singapore University of Technology and Design, Singapore,
%Email: \{zhang\_jun, yuenchau\}@sutd.edu.sg\\}
%
%\IEEEauthorblockA{$^{\dag}$ Institute of Communications Engineering, National Sun Yat-sen University\\
%Kaohsiung 804, Taiwan, Email: \{ckwen\}@ieee.org\\}
%
%\IEEEauthorblockA{$^{\ddag}$ National Mobile Communications Research Laboratory, Southeast University\\
%Nanjing 210096, P. R. China, Email: \{jinshi, xqgao\}@seu.edu.cn}\\
%}

\markboth{IEEE Wireless Communication Letters}%
{Yang \MakeLowercase{\textit{et al.}}:  NOVEL $5$G RaPro SYSTEM ARCHITECTURE}

\maketitle

\begin{abstract}
We propose a novel fifth-generation (5G) rapid prototyping (RaPro) system architecture by combining FPGA-privileged modules from a software defined radio (or FPGA-coprocessor) and high-level programming language for advanced algorithms from multi-core general purpose processors. The proposed system architecture exhibits excellent flexibility and scalability in the development of a 5G prototyping system. As a proof of concept, a multi-user full-dimension multiple-input and multiple-output system is established based on the proposed architecture. Experimental results demonstrate the superiority of the proposed architecture in large-scale antenna and wideband communication systems.
\end{abstract}

\begin{IEEEkeywords}
5G, prototype, system architecture, software defined radio, general purpose processor.
\end{IEEEkeywords}

\section{Introduction}
With emerging applications and services from wireless networks, the volume of mobile traffic and the number of
devices have exponentially increased. To accommodate these demands, researchers proposed several promising technologies,
such as massive multiple-input multiple-output (MIMO), millimeter wave (mmWave), and ultra-densification, for the development
of fifth-generation (5G) or next generation wireless communication systems. These technologies have been used
to address demands through different dimensions, such as spectrum efficiency enhancement, spectrum extension, and
network densification \cite{andrews2014will}.

To verify the anticipated gain from large-scale antenna array, researchers established several basic massive MIMO prototyping
systems. Based on their signal processing realization methods, the prototyping systems can be roughly classified into two categories, namely, FPGA-based prototyping system and general purpose processor(GPP)-based prototyping system. FPGA-based prototyping systems include Argos \cite{shepard2012argos,shepardthesis2012argos,shepard2013argosv2} developed by Rice University and LuMaMi \cite{vieira2014flexible} created by Lund University. Some examples of GPP-based prototyping systems are Open Air Interface (OAI)
massive MIMO testbed \cite{kaltenberger2015openairinterface,jiangopenairinterface} developed by Eurecom team, and BigStation \cite{yang2013bigstation} developed by Microsoft Research team with the Microsoft Research Software Radio (Sora) \cite{tan2011sora}. However, FPGA-based prototyping systems constantly suffer from long programming and compile time, whereas the performance of OAI and Sora are constrained by their specialized hardware driver and software stack.

%Compared with OAI massive MIMO testbed and BigStation, Argos and LuMaMi can satisfy the real-time requirements
%of larger antenna arrays (e.g., 100 antennas) and a wide transmission bandwidth (e.g., 20 MHz) with a longer development
%period as well as nontrivial conversion from a floating point computation to a fixed point computation. Meanwhile, OAI
%massive MIMO testbed and BigStation can provide priorities and flexibilities for innovative algorithm development because
%all baseband signals are processed on software. However, the GPP-based prototyping systems exhibit an inherent bandwidth
%limitation such as the software stack for OAI and the rigid system architecture for Sora, which may be insufficiently scalable for
%large-scale antenna systems and wideband mmWave systems.

\begin{figure}[tb]
\centering
\includegraphics[width=0.7\textwidth]{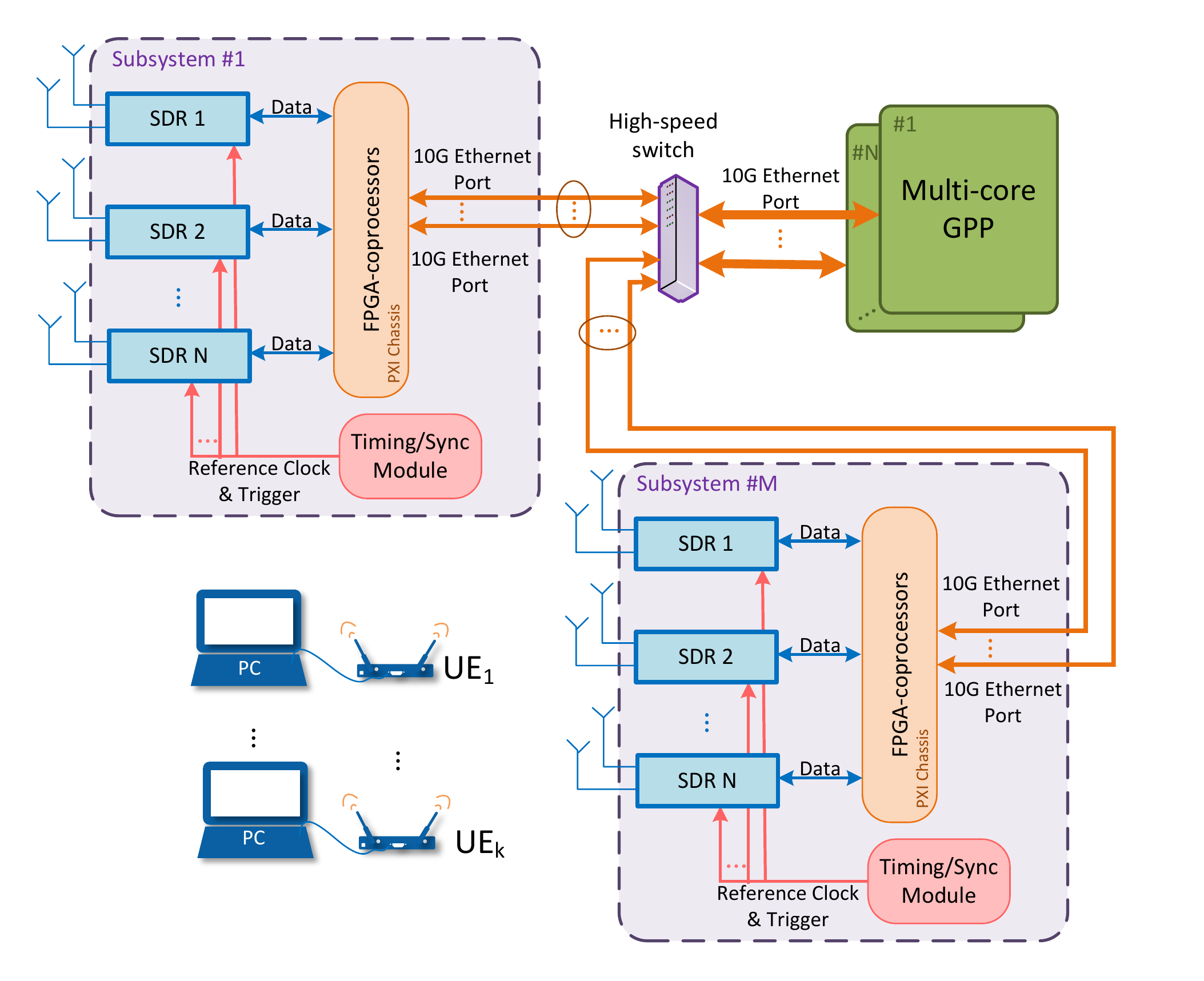} %0.45
\caption{Proposed system architecture comprising SDRs, FPGA-coprocessors, timing/synchronization module, high-speed 10G Ethernet switches, and multi-core GPPs.}
\label{fig:1}
\end{figure}

To address these limitations, we propose a novel system architecture by extracting the advantages of the software-defined radio (SDR) platforms and multi-core GPPs through a high-speed 10-Gigabit (G) Ethernet interface connection. In particular, we distribute FPGA-privileged modules into SDRs, such as FFT/IFFT; implement complex innovative algorithms, such as advanced receiver algorithms, in multi-core GPPs using high-level language; and connect SDRs and GPPs with 10-Gigabit Ethernet interfaces. The proposed system architecture is flexible and scalable in terms of the number of antennas and transmission bandwidth, and can be applied to various scenarios, such as full-dimension MIMO, centralized/distributed massive MIMO, mmWave, and cloud-based networking.
%The remainder of this letter is organized as follows. Section II provides a brief overview of our system architecture. Section III demonstrates an instance as a proof of concept by establishing a multi-user full-dimension MIMO prototyping system.

\section{System Architecture}
In this section, the proposed system architecture is briefly introduced, and the application scenarios are presented.
\vspace{-0.4cm}
\subsection{Overview}
Fig.~\ref{fig:1} illustrates the proposed system architecture, which comprises SDRs, FPGA-coprocessors, timing/synchronization
modules, high-speed 10G Ethernet switches, and multi-core GPPs. Compared with the conventional prototyping system architectures, our proposed system architecture introduces an FPGA-coprocessor and a high-speed Ethernet interface to enhance the capability of real-time signal processing and data transfer
between SDRs and GPPs.

SDRs contain RF chains provided with a unified reference clock and trigger signal by the timing/synchronization module in each subsystem. Data reordering and bandwidth splitting based on the antenna or subcarrier index are conducted in FPGA-coprocessors, which are assembled in a peripheral component interconnect eXtensions for instrumentation (PXI) chassis per subsystem. User datagram protocol (UDP)-packets and latency-constrained signal processing modules are implemented in the FPGA-coprocessors. High-speed switches aggregate or distribute data between subsystems and multi-core GPPs. Advanced baseband algorithms are obtained in multi-core GPPs through multiple parallel threads with high-level language in a Linux environment.
%When the user equipments transmit to the base station, the RF signals acquired by the antenna array are firstly conveyed to these SDRs through cables e.g. SMA cables. After down conversion, ADC, IQ imbalance correction, fractional decimation and other simply baseband preprocessing such as synchronization and FFT in SDRs, baseband data are transferred to FPGA-coprocessors for further processing e.g. bandwidth splitting, UDP packet formatting and transmission to 10GEthernet network ports. According to the labels contained in each UDP packet header, signal processing tasks can be arranged into multiple parallel threads in the alignment of frame schedule and subband partition. The downlink is in the inverse procedure of the uplink.

The proposed architecture offers the following advantages:

\begin{itemize}
  \item \emph{Scalability and flexibility.} The proposed system architecture is highly scalable in terms of the number of antennas
    and transmission bandwidth with the introduction of other subsystems, more subband (subframe) division, and
    other powerful multi-core GPPs, owing to the utilization of a subsystem and baseband data partition in a frequency
    (e.g., subband) or time (e.g., subframe) domain inside FPGA-coprocessors\footnote{The data transmission rate and peer-to-peer link pipes between FPGA-coprocessors are hardware-constrained.}. The system can also be conveniently
    configured as either centralized or distributed using multimode fibers for 10G Ethernet ports. Moreover, innovative
    algorithms can be deployed flexibly on either FPGA-coprocessors or GPPs.
  \item \emph{High computational capability.} The proposed architecture exhibits low-processing latency and high computation
    capability with the stack of FPGA-coprocessors and advanced GPPs. This is because multiple parallel threads are used
    in multi-core GPPs, in which each thread is allocated to an individual core (running at 2.8\,GHz versus 200\,MHz in FPGA-coprocessors), and each core is obliged to finish small detailed tasks in a parallel or pipeline manner.
%    Therefore, a single operation (e.g., ADD operation) takes 0.36ns for processing, which indicates that lower processing latency can be achieved in advanced GPPs.
  \item \emph{Rapid and easy development.} Innovative algorithms and large dimension matrix computation can be developed
    on multi-core GPPs by programming with high-level language, such as C/C++, in conjunction with Intel Math
    Kernel Library (Intel MKL) \cite{IntelMKL}, which is a highly optimized and commonly used math library for processors.
    Algorithm development with this approach is much more rapid and convenient than with Verilog for FPGAs in
    conventional prototyping systems.
\end{itemize}
\vspace{-0.4cm}
\subsection{Application Scenarios}
The proposed system architecture can be used for various scenarios, such as full-dimension MIMO, centralized/distributed massive MIMO, mmWave, and cloud-based networking.

\textbf{Centralized/distributed massive MIMO:} Excluding the 10G Ethernet connection and multi-core GPPs, our earlier work \cite{yang2016design} revealed a 128-antenna massive MIMO prototyping system similar to the proposed system architecture in Fig.~\ref{fig:1}. Therefore, the proposed system architecture is made feasible by only verifying the data transmission rate from FPGA-coprocessors to multi-core GPPs and the real-time computation capability in multi-core GPPs, as demonstrated in Section III.
%In addition, besides the centralized massive MIMO, distributed massive MIMO prototyping system can still be achieved by the usage of long fibers connecting between subsystem and multi-core GPPs.

\textbf{mmWave:} Wide bandwidth, high sample rate, and high data throughput are among the main challenges in prototyping
an mmWave system. An mmWave prototyping system can be established based on the usage of data partition in frequency or time domain in FPGA-coprocessors with the aid of more high-speed Ethernet switches and more advanced multi-core GPPs, in which each GPP is responsible for subband (or subframe) data processing.

\textbf{Cloud-based network:} A prototyping system based on the proposed system architecture can inherently exchange data
with an external network using a UDP packet and an Ethernet interface. Therefore, a cloud-based network with a
remote server cluster can be conveniently constructed, and this network is suitable for a wide area spectrum cognition or big
data collection for machine learning.

\begin{figure*}[htb]
\centering
\includegraphics[width=1.0\textwidth]{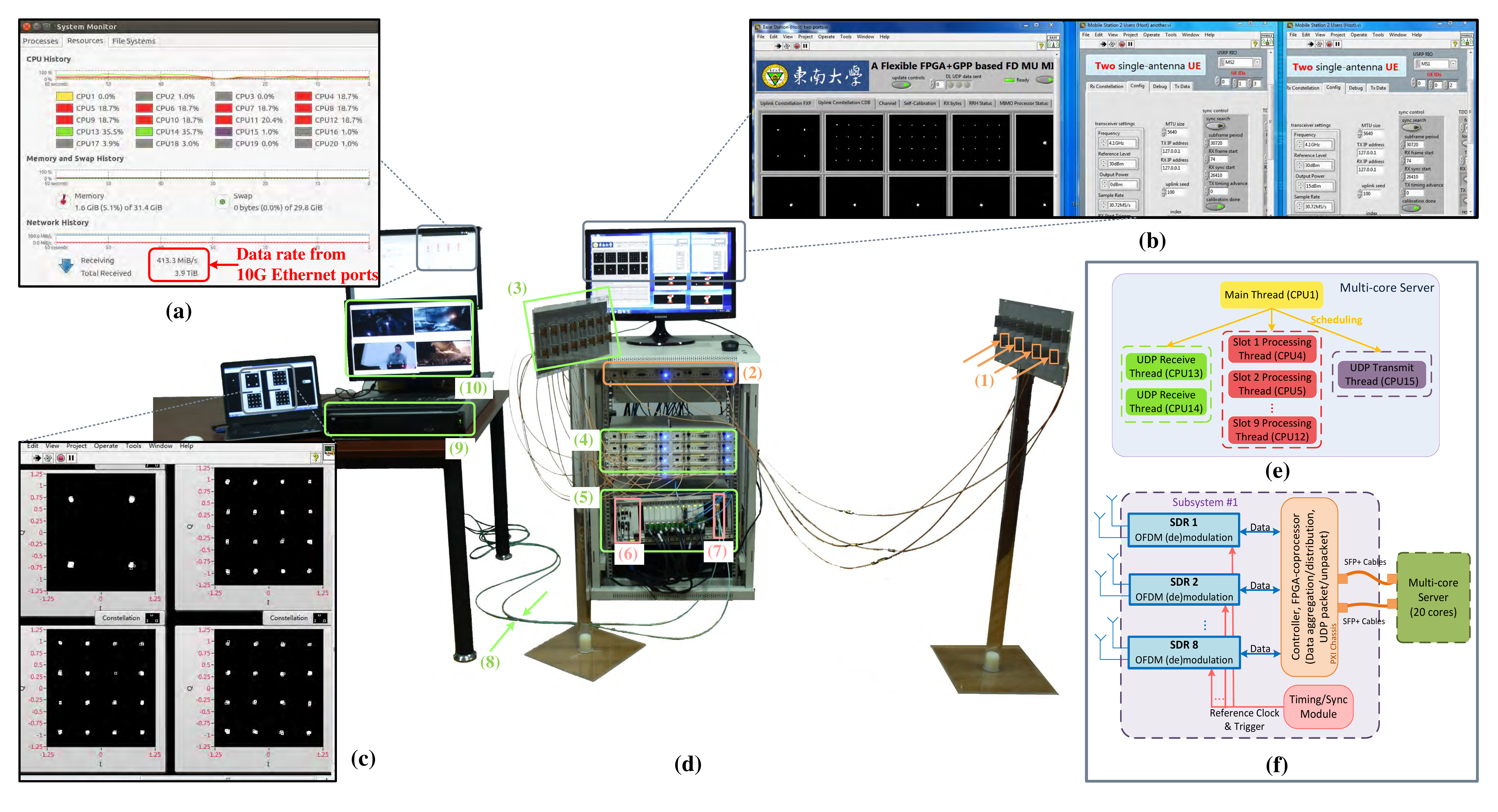} %0.9
\caption{Assembled multi-user full-dimension MIMO communication system based on the proposed system architecture. (a) System monitor for the server in a
Linux environment with the duty cycles of CPU cores, memory usage, and real-time data transmission rate of the server's 10G Ethernet ports. (b) Configuration and debugging program for SDR nodes (both BS and UEs) on the controller and UEs can be implemented in individual PCs, that is, UEs become separated with BS through synchronization over the air. (c) Constellations for four users are demodulated by the server and conveyed to the video client (a laptop) via a UDP port. (d) The assembled practical system composed of 1) four individual dipole antennas for four single-antenna users, 2) two SDR nodes emulating four single-antenna users, 3) an $8 \times 2$ uniform planar array, 4) eight SDR nodes emulating a sixteen-antenna base station, 5) a PXI chassis connecting the SDR nodes and FPGA-coprocessor, 6) a controller responsible for the configuration and initialization of SDR nodes, 7) an FPGA-coprocessor configured with four 10G Ethernet ports, 8) SFP+ cables, 9) a multi-core server, and 10) four recovered videos by the server for four users on a video client. (e) Parallel multi-thread processing procedure in the server corresponding to the system monitor in (a). (f) System architecture relevant to (d).}
\label{fig:2}
\end{figure*}

\section{A Full-Dimension MIMO system}
In this section, a constructive example is provided to validate the feasibility and flexibility of our proposed system
architecture by establishing a multi-user full-dimension MIMO prototyping system.
\vspace{-0.4cm}
\subsection{Signal Model}
Considering a multi-user full-dimension MIMO communication system, where $K$ single-antenna user equipments (UEs) communicate with a base station (BS) having an $L_1 \times L_2$ uniform planar antenna array, we apply an orthogonal frequency division multiplexing (OFDM) technology. Different users transmit frequency orthogonal pilots in the uplink training phase. Additionally, least square (LS) channel estimation and linear minimum mean square error (LMMSE) detector are employed in the base station. The transmission procedure is divided into two phases, namely, the training phase and data transmission phase, which are consistent with those used in our past work \cite{yang2016design}.

\vspace{-0.4cm}
\subsection{Prototype Setup}
To implement the multi-user full-dimension MIMO communication system with the proposed system architecture, we employ eight SDR nodes of the USRP RIO series manufactured by National Instruments. Each SDR node consists of two RF transceivers of 120\,MHz bandwidth and a programmable FPGA responsible for distributed signal processing, such as reciprocity calibration or OFDM (de)modulation. Fig. \ref{fig:2}(d) shows the assembled multi-user
full-dimension MIMO communication system comprising a clock distribution module, an FPGA-coprocessor card configured with four 10G Ethernet ports, a high-data throughput PXI chassis, a server containing 20 Intel Xeon E5-2680 v2 @2.8\,GHz processors, and two small form-factor pluggable plus (SFP+) cables.

Similar to the frame structure described in our previous study \cite{yang2016design}, a simplified LTE-like 10\,ms radio frame structure is used. In this structure, Subframe\,0 is used for synchronization and Subframes\,1-9 are used for uplink and downlink data transmission. Once a successful synchronization occurs between BS and UEs, frequency orthogonal uplink pilot OFDM symbols followed by uplink data OFDM symbols are transmitted from four single-antenna users. The RF signals acquired by the $8 \times 2$ antenna array are initially delivered to eight SDR nodes for down conversion and OFDM
demodulation. The obtained baseband data are subsequently converged in the FPGA-coprocessor through the PXI chassis for bandwidth splitting and UDP packaging. Then, UDP-packaged data are conveyed to the server by two 10G Ethernet ports and two SFP+ cables.

According to the labels contained in each UDP packet header, signal processing tasks in a server are arranged to nine threads in parallel, and each thread performs the task of one subframe data processing. The signal processing tasks include the channel estimation and MIMO detection. In Fig.~\ref{fig:2}(e), the following 13 parallel threads are found in the server for our 16-antenna system: one main thread (represented by yellow), two UDP receive threads (represented by green), nine subframe processing threads (represented by red), and one UDP  transmission thread (represented by purple). Each thread is
allocated with an individual CPU core. Thus, 13 of the 20 CPU cores in the server are leveraged and displayed by the system monitor in Fig.~\ref{fig:2}(a).

\begin{figure}[htb]
\centering
\includegraphics[width=0.6\textwidth]{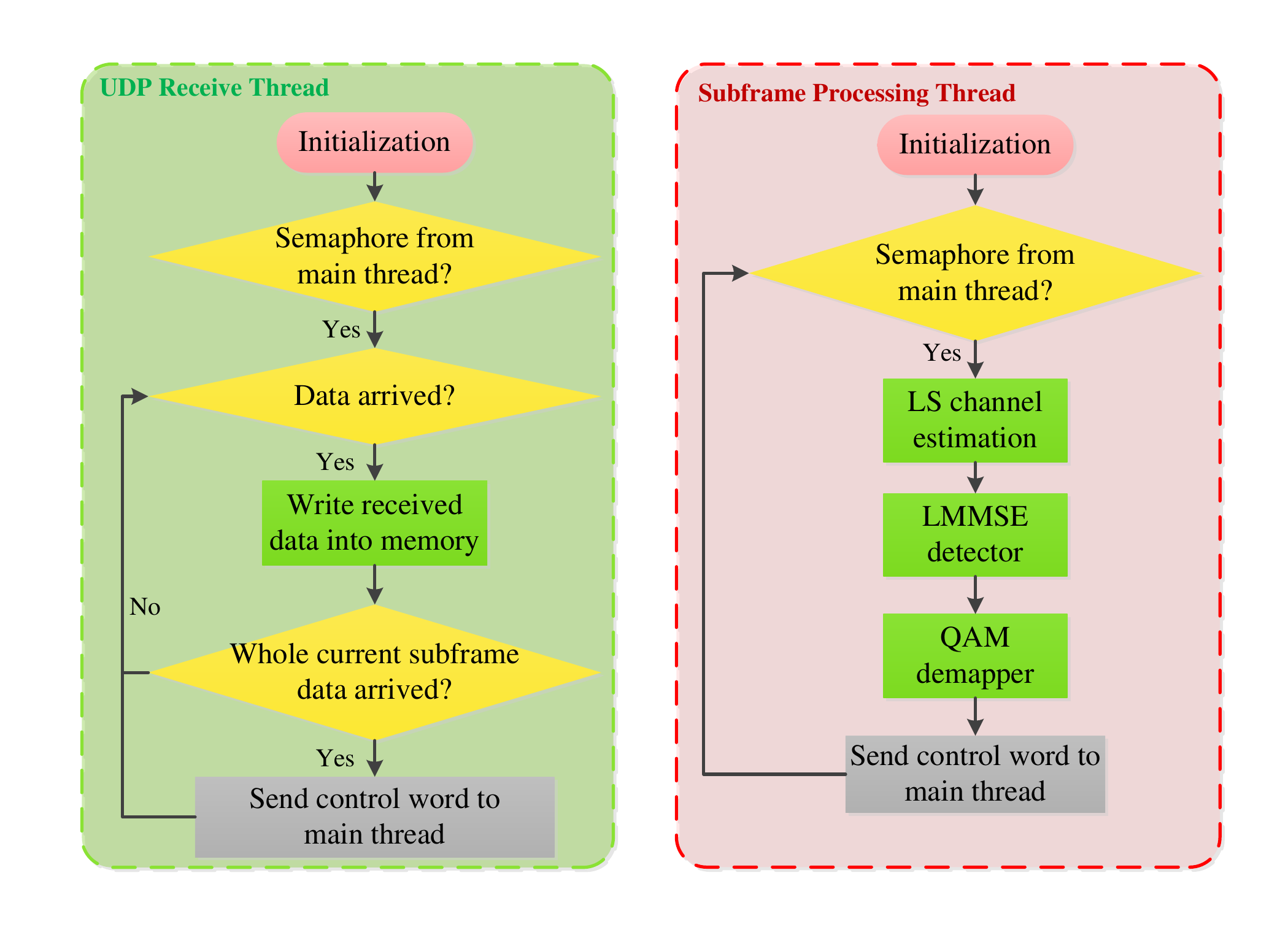} %0.43
\caption{Detailed processing procedure of the UDP receive thread and the subframe processing thread. After initializing the system, the UDP receive
thread waits for the semaphore from the main thread. Once the semaphore arrives, the UDP receive thread begins to poll the 10G Ethernet port and writes
the received data to memory. When the entire subframe data are received, a related control word is sent to the main thread. According to the received
control word, the main thread immediately sends the semaphore to one of the processing threads in Subframes\,1-9 [Fig. \ref{fig:2}(e)]. When the subframe processing thread receives the semaphore from the main thread, data are read from the specified memory address; then signal processing such as LS channel estimation or LMMSE detection instantly begins. Demodulated data are written back to memory. As the current subframe processing is completed, the subframe
processing thread informs the main thread that the demodulated data must be delivered to the video client for constellation and video display. The main
thread subsequently schedules the UDP transmit thread to finish this task. These processes are implemented in C/C++ with Intel MKL.}
\label{fig:3}
\end{figure}

Two UDP receive threads collect data from the FPGA-coprocessor via two 10G Ethernet ports. Nine subframe processing threads are used for nine subframe data processing, and one UDP transmit thread is utilized for demodulated data transfer from the server to the video client for the constellation and video display. The main thread is utilized to schedule the other threads. Moreover, the procedures of the nine subframe processing threads are consistent, except the data to be processed are from different subframes with respect to that task arrangement. Specifically, the data of Subframes\,1 and 2 are
processed in the Subframe 1 and the Subframe 2 Processing Threads, respectively. The same process occurs in the other subframes. The detailed processing procedures of each subframe processing thread and the UDP receive thread are illustrated in Fig.~\ref{fig:3}.

\vspace{-0.4cm}
\subsection{Experiment Results}

To validate the feasibility and flexibility of the proposed system architecture, we provide our experimental results in terms of data transmission, CPU load analysis, and data streaming analysis based on proposed multi-user full-dimension system.

%\begin{enumerate}[1)]
\textbf{Data transmission.} In the prototype system [Figs. \ref{fig:2}(b)-(d)], the carrier frequency is 4.1\,GHz, the sample rate is 30.72\,MS/s, and the data transmission bandwidth is 20\,MHz. QPSK is adopted by user 0, and the three other users use 16-QAM. Under the current settings, video transmission is
successful. We also test 64-QAM and 256-QAM for data transmission. With the assistance of channel estimate interpolation, we find that 64-QAM can successfully perform video transmission while much mosaic appears in 256-QAM. This finding indicates that the channel coding is necessary.

\textbf{CPU load analysis.} In implementing the algorithm in multi-core GPP, the CPU load representing the duty cycle of CPU cores requires analysis. The calculated CPU cycles and theoretical duty cycle under different scenarios are listed in Table \ref{tab:cpu cycles}. Approximately $4.806 \times 10^6$ cycles or $\frac{{4.806 \times {{10}^6}}}{{2.8 \times {{10}^9}}} = 1.716$\,ms are spent by each core to process the signals of the
16-antenna system with four single-antenna users. The theoretical duty cycle is 17.16\% based on the 10\,ms radio frame structure. This theoretical value nearly matches the practical duty cycle of 18.7\% [Fig. \ref{fig:2}(a)] when some trivial operations are disregarded. Table \ref{tab:cpu cycles} also indicates that the computation capability of the server is sufficient to handle 12 users regardless of whether a 16- or 128-antenna system is used. Clearly, a wideband or a large-scale antenna prototyping system with more advanced CPU cores can be applied.

\textbf{Data streaming analysis.} In configuring one uplink pilot symbol and two uplink data symbols per slot (except for Subframe\,0) with 16-bit (or 2-byte) analog-to-digital converter (ADC), the theoretical data rate of two 10G Ethernet ports is $30.72({\rm{MS/s}}) \times \frac{{1200 \times 54}}{{307200}} \times 2({\rm{I,Q}}) \times 2({\rm{bytes}})\times16({\rm{antennas}})=414.72{\rm{MB/s}}$, nearly matching the practical data rate $413.3{\rm{MB/s}}$ displayed in Fig. \ref{fig:2}(a). Hence, each port is approximately $414.72{\rm{MByte/s}} \times 8 /2\approx1.7$\,Gbps $<$ 10\,Gbps, which is the line rate of 10G Ethernet port and indicates the potential of a 128-antenna system utilizing four 10G Ethernet ports individually running at 6.6\,Gbps. Increased data throughput, widened bandwidth, and additional antennas can also be supported by maximizing FPGA-coprocessor cards and transplanting channel estimation
into FPGA.

\textbf{Superiority analysis.} Table \ref{tab:comparison} compares existing prototyping systems. Owing to the short compiling time (e.g., several seconds in GPP) and the unnecessary conversion from floating-point calculation to fixed-point calculation, the comparison of Argos and LuMaMi shows that RaPro can satisfy the real-time requirements of large antenna arrays and a wide transmission bandwidth with a significantly shorter development period. Meanwhile, RaPro breaks the performance bottleneck in the BigStation and OAI testbeds using a customized software architecture in GPPs while maintaining priorities and flexibilities in developing innovative algorithm.

%Compared with OAI massive MIMO testbed and BigStation, Argos and LuMaMi can satisfy the real-time requirements
%of larger antenna arrays (e.g., 100 antennas) and a wide transmission bandwidth (e.g., 20 MHz) with a longer development
%period as well as nontrivial conversion from a floating point computation to a fixed point computation. Meanwhile, OAI
%massive MIMO testbed and BigStation can provide priorities and flexibilities for innovative algorithm development because
%all baseband signals are processed on software. However, the GPP-based prototyping systems exhibit an inherent bandwidth
%limitation such as the software stack for OAI and the rigid system architecture for Sora, which may be insufficiently scalable for
%large-scale antenna systems and wideband mmWave systems.

\begin{table}[tb]
\caption{CPU load analysis for subframe processing CPU cores under different settings.}\label{tab:cpu cycles}
\centering
\scriptsize
  \begin{tabular}{ccccccc}
    \toprule
     \multirow{2}*{Settings} & \multicolumn{3}{c}{Module (CPU cycles)} & \multirow{2}*{Total cycles} &\multirow{2}*{Time} &\multirow{2}*{Duty cycle(T)} \\
     \cmidrule{2-4}
     (20MHz) &  $\bf H_{est}$ &  Detector &  Others & (Per core) & @2.8GHz & (Per core)\\
     \midrule
     \rowcolor{mygray}
     16x4 & 66k & 3800k & 940k & 4806k & 1.716ms & 17.16\% \\
     16x12 & 188k & 5000k & 1340k & 6128k & 2.189ms & 21.89\% \\
    \rowcolor{mygray}
     128x12 & 1504k & 15240k & 6520k & 23264k & 8.309ms & 83.09\% \\
    \bottomrule
  \end{tabular}
\end{table}

\begin{table}[tb]
\caption{Comparison among existing prototyping systems.}\label{tab:comparison}
\vspace{-0.4cm}
\scriptsize
\begin{center}
\begin{tabu} to 0.68\textwidth{X[2.5,c] X[2,c] X[2,c] X[c] X[2,c] X[1.5,c] X[2,c]} %0.53
    \toprule
     & \tabincell{c}{\# of BS\\ antennas} & Bandwidth & \tabincell{c}{FFT\\ Size} & Hardware & \tabincell{c}{Develop.\\ period} & \tabincell{c}{Scalability/\\Flexibility} \\
     \midrule
     \rowcolor{mygray}
     RaPro & up to 128 & 20MHz & 2048 & GPP+FPGA & Short & High/High \\
     \cite{yang2016design} & 128 & 20MHz & 2048 & FPGA & Long & High/Low \\
    \rowcolor{mygray}
     Argos & up to 96 & 625KHz & - & FPGA & Long & Med./Low \\
     LuMaMi & 100 & 20MHz & 2048 & FPGA & Long & High/Low\\
     \rowcolor{mygray}
     BigStation & 12 & 20MHz & 64 & GPP+Sora & Short & Low/High \\
     OAI testbed & up to 64 & 5MHz & 512 & OAI+FPGA & Short & Med./Med. \\
    \bottomrule
\end{tabu}
\end{center}
\end{table}
%  \begin{equation}\label{eq:data rate}
%  \scriptsize
%    30.72 \times \frac{{1200 \times 54}}{{307200}} \times 4 \times 16 = 414.72MB/s \approx 413.3MB/s
%  \end{equation}

%\end{enumerate}

\section{Conclusion}

%A novel flexible 5G prototyping system architecture is proposed by using an SDR platform and multi-core GPPs with high-speed 10G Ethernet interface connection. Experiment results reveal that the proposed system architecture is advantageous for advanced algorithm implementation and thus
%can be used for many applications.
A novel flexible 5G prototyping system architecture is proposed using an SDR platform and multi-core GPPs with high-speed 10G Ethernet interface connection. Experiment results reveal that multi-core GPPs are advantageous for rapid advanced algorithm implementation, and that the 10G Ethernet interface connection creates a scalable and flexible system in terms of antennas and transmission bandwidth. The proposed architecture is applicable to various applications.

%\section*{Acknowledgment}
%This work was supported by National Natural Science Foundation of China under Grants 60925004, 60902009 and 61101089, and National Science and Technology Major Project of China under Grants 2011ZX03003-001 and 2011ZX03003-003-03.

%\bibliographystyle{IEEEtran}
%\bibliography{References_2017_1}

\end{document}